\documentclass[conference]{IEEEtran}
\IEEEoverridecommandlockouts
\usepackage{cite}
\usepackage{amsmath,amssymb}
\usepackage{graphicx}
\usepackage{booktabs}
\usepackage{array}
\usepackage{url}
\usepackage{multirow}
\usepackage{stfloats}
\usepackage{balance}
\usepackage{xcolor}
\usepackage{algorithm}
\usepackage{algpseudocode}

\begin{document}

\title{PMR-Aided Shifted-Pruning SCL-Flip Decoding for CRC-Aided Polar Codes}

\author{
\resizebox{1\textwidth}{!}{%
\begin{tabular}{@{}c@{}}
Niloufar~Hosseinzadeh$^{*}$,
Mohammad~Morsali$^{\dagger}$,
Mohsen~Moradi$^{\ddagger}$,
and Hessam~Mahdavifar$^{\S}$\\[3pt]
$^{*}$Department of Electrical and Computer Engineering, Korea University, Seoul 02841, Republic of Korea\\
$^{\dagger}$ School of Telecommunications and Aerospace Engineering , Universitat Polit\`ecnica de Catalunya, Castelldefels 08860, Spain\\
$^{\ddagger}$School of Electrical, Computer and Energy Engineering, Arizona State University, Tempe, AZ 85287, USA\\
$^{\S}$Department of Electrical \& Computer Engineering, Northeastern University, Boston, MA 02115, USA\\[3pt]
Emails: niloufarhz@korea.ac.kr$^{*}$,
mohammad.morsali@estudiantat.upc.edu$^{\dagger}$,
mmorad11@asu.edu$^{\ddagger}$,
h.mahdavifar@northeastern.edu$^{\S}$
\end{tabular}%
}
}

\maketitle

\thispagestyle{plain}
\pagestyle{plain}

\begin{abstract}
We propose a posterior-mass rescue (PMR)-aided shifted-pruning
SCL-flip decoder for CRC-aided polar codes. After CRC-aided SCL
(CA-SCL) and differential-metric shifted-pruning recovery fail
to find a CRC-valid candidate, PMR locally continues discarded
prefixes stored during the initial CA-SCL pass, ranked by
estimated posterior mass, without increasing the shifted-pruning
retry budget.
Simulation results for a $(128,64)$ code with the Reed--Muller (RM) rate
profile and a $(256,128)$ code with the polarization-weight (PW) rate profile,
both using an 11-bit CRC, show lower observed FER than SCLF-Diff
$(T=15)$ and the considered literature baselines under the evaluated
recovery budgets, with observed FER close to that of SCLF-Diff $(T=20)$.
At the reported moderate and high SNRs, the average reconstruction-aware
path-step count is within a few percent of that of SCLF-Diff $(T=20)$,
while at $2.0$ dB it is $9.4\%$ and $4.3\%$ higher for the two codes,
respectively. The proposed decoder also has a larger worst-case
path-step bound.
\end{abstract}


\section{Introduction}
Polar codes are the first class of explicitly constructed codes proven to
achieve the capacity of binary-input memoryless symmetric channels under
successive cancellation (SC) decoding~\cite{arikan2009}.
At short-to-moderate block lengths, successive-cancellation list (SCL)
decoding improves error-correction performance by maintaining multiple
candidate paths. A cyclic redundancy check (CRC) further assists final
selection, with the lowest-metric CRC-valid candidate chosen from the
list~\cite{tal2015}. Nevertheless, CRC-aided SCL (CA-SCL) decoding can fail
when intermediate pruning removes the correct path, leaving no correct
candidate in the final list.

Successive-cancellation flip (SCF) decoding addresses detected failures
through additional decoding attempts that reverse selected unreliable
decisions~\cite{afisiadis2014scf}. Its list-based counterpart, SCL-flip
(SCLF), performs CRC-triggered retries that alter selected decisions or
pruning operations~\cite{yu2018sclbf,cheng2019bitflip}. In shifted-pruning SCLF (SP-SCLF), an
additional trial shifts the surviving-path window to retain candidates that
ordinary SCL pruning would discard~\cite{rowshan2022sclistflip}. Recovery
effectiveness depends on the reliability of the retry-position ranking and
the available retry budget. Increasing that budget can reduce frame-error
rate (FER), but requires additional path expansions and increases decoding
latency. Consequently, practical recovery design must jointly consider
which failed decisions to revisit and how much computation each additional
decoding attempt requires. Generalized SCF restarts~\cite{sagitov2025grm} and limited-location SCLF
restarts~\cite{pillet2025restart} reuse computations from earlier attempts
without changing error-correction performance. Partitioned SCLF combines
dynamic flip metrics, multiple path flips, early termination, and
partition-based restarts~\cite{pillet2025partitioned}.
Threshold-learning-based SCLF uses a dual-threshold strategy and segmented
CRC to reduce redundant retries and permit early
termination~\cite{sun2026tssclf}. Multi-attempt SCL reproduces standard SCL
error-correction performance with lower average
complexity~\cite{yuan2026mascl}. Complementarily, soft-output SCL approximates probability mass
in unvisited subtrees under uniform unexplored
completions~\cite{yuan2025soscl}. Successive-cancellation
ordered-search (SCOS) decoding ranks alternative branches and
progressively updates its search list during partial SC
re-decoding~\cite{yuan2024scos}.

We propose a posterior-mass rescue (PMR)-aided
shifted-pruning SCLF decoder that locally continues prefixes
discarded exclusively during the initial CA-SCL pass, using a
bounded bank, coverage-based selection, and prefix-specific
lists. Unlike stack decoding~\cite{niu2012stack} and
SCOS~\cite{yuan2024scos}, which perform ordered tree search
during decoding, PMR is activated only after CA-SCL and
differential-metric SCLF recovery~\cite{ivanov2021gsclf}
fail to find a CRC-valid candidate. Unlike
SO-SCL~\cite{yuan2025soscl}, PMR uses the subtree-mass
approximation to control cross-depth ranking and bounded
local continuation. The contributions are:

\begin{itemize}
    \item A PMR-aided shifted-pruning SCLF decoder is proposed for CRC-aided
    polar codes, where PMR is activated only after both CA-SCL and the
    differential-metric SCLF recovery stage fail.

    \item A bounded discarded-prefix bank is built during the initial
    CA-SCL pass. PMR repurposes the subtree-mass approximation
    of~\cite{yuan2025soscl} for cross-depth prefix ranking,
    coverage-based selection, and local-list allocation under a
    bounded budget.

    \item An empirical evaluation quantifies the FER benefit and
    additional reconstruction-aware path-step cost of adding PMR to
    SCLF-Diff $(T=15)$. Comparisons with SCLF-Diff $(T=20)$ and
    PMR-only variants characterize the configured recovery strategies,
    while a score ablation evaluates the frozen-suffix correction
    under common recovery parameters in separate numerical results. 
    
\end{itemize}

\section{System Model}
\label{sec:system_model}

We consider a CRC-aided polar-coded transmission over a binary-input
additive white Gaussian noise (BI-AWGN) channel. Let $N=2^n$ denote the
polar-code block length, $A$ the data length, and $r$ the CRC length.
Appending the CRC produces $K=A+r$ polar information bits, which are placed
at the indices in the information set $\mathcal{I}$, with
$|\mathcal{I}|=K$. The remaining indices form the frozen set
$\mathcal{F}=\{0,\ldots,N-1\}\setminus\mathcal{I}$ and are set to zero.

Using the standard polar transform~\cite{arikan2009}, the input vector
$\mathbf{u}\in\{0,1\}^{N}$ is encoded as
$\mathbf{x}=\mathbf{u}\mathbf{G}_{N}$ over $\mathrm{GF}(2)$, where
$\mathbf{G}_{N}=\mathbf{F}^{\otimes n}$ is the Ar{\i}kan polar transform.
The encoded bits are BPSK-modulated and transmitted as
$1-2x_i$ over the AWGN channel, yielding
$y_i=1-2x_i+n_i$, where $n_i\sim\mathcal{N}(0,\sigma^2)$.
Accordingly, the channel LLR supplied to the decoder is
$L_i=2y_i/\sigma^2$. The information set $\mathcal{I}$ is selected
according to the rate profile specified for each simulated code in
Section~\ref{sec:simulation_setup}.

\section{Proposed Method}
\label{sec:proposed_method}

\subsection{CA-SCL Foundation}
\label{subsec:ca_scl_foundation}

The standard CA-SCL decoder maintains at most $L$ candidate paths, where $L$ is the list size. At an
information position $i\in\mathcal{I}$, each surviving path $\ell$ is extended
by $b\in\{0,1\}$ using the LLR-domain path-metric update~\cite{balatsoukas2015llr}
\begin{equation}
\mathrm{PM}_{\ell,b}^{(i)}
=
\mathrm{PM}_{\ell}^{(i-1)}
+
\log\!\left(1+\exp\left[-(1-2b)L_{\ell}^{(i)}\right]\right),
\label{eq:path_metric}
\end{equation}
where $L_{\ell}^{(i)}$ is the path-dependent LLR and
$\mathrm{PM}_{\ell,b}^{(i)}$ is the metric of path $\ell$ extended
by $b$. This exact metric preserves the original Tal--Vardy
likelihood-based path ordering~\cite{tal2015,balatsoukas2015llr}.
Retained paths are reindexed, with metrics
$\mathrm{PM}_{\ell}^{(i)}$.
Metrics start at zero and use~\eqref{eq:path_metric}
at every position, with $b=0$ for frozen bits; stored metrics
share a cumulative scale across depths.

At the end of decoding, CA-SCL returns the lowest-metric CRC-valid candidate
in the final list $\mathcal{A}_{N-1}$, denoted by
$\hat{\mathbf{u}}_{\mathrm{CA}}$, where $\mathcal{A}_i$ denotes the set of
at most $L$ paths surviving after position $i$. If the final list contains no CRC-valid candidate, the shifted-pruning recovery
stage is activated.

\subsection{Shifted-Pruning Recovery Stage}
\label{subsec:sp_sclf_stage}

Following a CA-SCL failure, we employ a differential-metric SCLF recovery
stage based on the critical-set~\cite{ivanov2021gsclf}. At a full-list information position $i$, the
$2L$ candidates are ordered by increasing metric as
$c_i^{(0)},c_i^{(1)},\ldots,c_i^{(2L-1)}$. The pruning positions are ranked
using the gap between the first discarded candidate and the best candidate:
\begin{equation}
\Delta_i^{\mathrm{SP}}
=
\mathrm{PM}_{c_i^{(L)}}^{(i)}
-
\mathrm{PM}_{c_i^{(0)}}^{(i)},
\label{eq:sp_sclf_score}
\end{equation}
where a smaller value indicates that the first discarded path is closer
in path metric to the best candidate.

The recovery stage selects up to $T$ positions with the smallest scores.
The selected positions are processed in increasing order of
$\Delta_i^{\mathrm{SP}}$. At each selected position $\tau$, an additional
shifted-pruning trial retains the window
$c_\tau^{(k)},\ldots,c_\tau^{(k+L-1)}$
instead of the standard window
$c_\tau^{(0)},\ldots,c_\tau^{(L-1)}$,
where $k$ denotes the shift of the surviving path window,
after which ordinary SCL decoding continues.

The proposed decoder uses $k=L$, so the trial retains the $L$ paths
discarded by conventional pruning, consistent with the symbol-wise SCLF
operation in~\cite{ivanov2021gsclf}. For comparison, SP-SCLF uses the pruning-position metric in~\cite[Eq.~(10)]{rowshan2022sclistflip}, with $k=L/2$ and $T=15$.

Decoding stops at the first shifted-pruning trial whose final list contains
a CRC-valid candidate, and the lowest-metric CRC-valid path in that trial
is returned as $\hat{\mathbf{u}}_{\mathrm{SP}}$. If none of the $T$ trials
succeeds, PMR is activated.

\subsection{PMR Fallback}
\label{subsec:pmr_fallback}

During the initial CA-SCL pass, PMR stores discarded prefixes
in a bank indexed by $\mathcal{B}$, with
$|\mathcal{B}|\le B_{\max}$. Prefix $j\in\mathcal{B}$ contains
the decisions through its pruning position $i_j$; its position
and metric $\mathrm{PM}_{j}^{(i_j)}$ are also stored.
The bank is used only if CA-SCL and shifted-pruning recovery
both fail to find a CRC-valid candidate. Path-specific LLR and
partial-sum states are not retained; they are reconstructed
from channel LLRs and stored decisions before suffix
continuation, avoiding complete state storage per entry.
Section~\ref{sec:simulation_setup} includes this reconstruction
cost.

To rank prefixes discarded at different depths, PMR adopts the
unvisited-subtree mass approximation used in
SO-SCL~\cite{yuan2025soscl}. Under the auxiliary distribution used
by SC decoding, $\exp(-\mathrm{PM}_{j}^{(i_j)})$ represents the
mass of prefix $j$ before enforcing future constraints.
Approximating unexplored completions as uniform introduces a
factor $1/2$ for each remaining frozen position. Let
\begin{equation}
F_{i_j}
=
\left|\{t\in\mathcal{F}:t>i_j\}\right|
\label{eq:future_frozen}
\end{equation}
denote the number of frozen positions remaining after $i_j$.
The approximate completion mass is then
\begin{equation}
\widehat{M}_j
\propto
\exp\left(-\mathrm{PM}_{j}^{(i_j)}\right)
2^{-F_{i_j}}.
\label{eq:approx_prefix_mass}
\end{equation}
Taking the negative logarithm and omitting a prefix-independent
normalization constant gives the ranking score
\begin{equation}
R_j
=
\mathrm{PM}_{j}^{(i_j)}
+
F_{i_j}\log 2.
\label{eq:pmr_rank}
\end{equation}
This approximation ignores the dependence of future frozen-bit
probabilities on the prefix and channel realization; future CRC
constraints are also excluded from the score. Thus, $R_j$ is a
ranking heuristic, with lower values indicating more promising
discarded prefixes.

When more than $B_{\max}$ discarded prefixes are available, the bank keeps only
the lowest-score prefixes according to \eqref{eq:pmr_rank}. An estimated
posterior-mass weight is then assigned to each stored prefix:
\begin{equation}
w_j
=
\frac{\exp(-R_j)}
{\sum_{q\in\mathcal{B}}\exp(-R_q)} ,
\qquad j\in\mathcal{B}.
\label{eq:pmr_weight}
\end{equation}
Because $\mathcal{B}$ contains at most $B_{\max}$ retained prefixes,
$w_j$ is an approximate within-bank posterior-mass weight. The
normalization $\sum_{j\in\mathcal{B}}w_j=1$ is conditional on the prefix
being present in the truncated bank and does not account for discarded
prefixes excluded from $\mathcal{B}$. Consequently, $\rho=0.99$ denotes
$99\%$ of the approximate mass within the retained bank, rather than
$99\%$ of the total posterior mass of all paths discarded during SCL
decoding. Prefixes with larger $w_j$ are treated as more promising
within the retained bank.
The prefixes are sorted in descending order of $w_j$. Let
$j_1,j_2,\ldots$ denote the sorted prefix indices, where
$w_{j_1}\ge w_{j_2}\ge\cdots$. The PMR recovery stage selects a prefix set
whose cumulative approximate within-bank posterior mass is at least the
coverage target $\rho$, while also enforcing a minimum selected-prefix count $M_{\min}$. The
number of selected prefixes is
\begin{equation}
m
=
\min\left\{
|\mathcal{B}|,
\;
\max\left\{
M_{\min},
\;
\min\left\{s:\sum_{q=1}^{s} w_{j_q}\ge \rho\right\}
\right\}
\right\}.
\label{eq:coverage_selection}
\end{equation}
The selected prefix set is then
$\mathcal{S}_{\rho}=\left\{j_1,\ldots,j_m\right\}$,
which is used for local continuation in the PMR fallback stage.

\subsection{Local Continuation from Stored Prefixes}
\label{subsec:local_continuation}

For a selected prefix $j$ discarded at $i_j$, PMR reconstructs its
path-specific SC state by replaying the stored decisions
$\hat{\mathbf{u}}_j^{0:i_j}$ and then decodes positions
$i_j+1,\ldots,N-1$. Reconstruction is performed only for prefixes processed
before CRC-based early stopping. The local decoder is initialized with
$\mathrm{PM}_j^{(i_j)}$ and uses a prefix-specific list size
$L_j^{\mathrm{loc}}$. If the prefix bank is empty, PMR is skipped.
Otherwise, $Q\geq|\mathcal{S}_{\rho}|$ is required
to allocate at least one local path per selected prefix. Under the total local-list budget $Q$, the allocation is
\begin{equation}
L_j^{\mathrm{loc}}
=
\min\left\{
L_{\max}^{\mathrm{loc}},
1+
\left\lfloor
(Q-|\mathcal{S}_{\rho}|)
\frac{w_j}{\sum_{q\in\mathcal{S}_{\rho}}w_q}
\right\rfloor
\right\},
\quad j\in\mathcal{S}_{\rho}.
\label{eq:local_list_allocation}
\end{equation}
Thus, more probable prefixes receive larger local lists while the total
recovery budget remains bounded. Selected prefixes are processed in
decreasing order of $w_j$ in groups of at most $R$ prefixes. After all
continuations in a group have been checked, PMR stops processing further
groups for a frame if a CRC-valid candidate has been found and returns the
lowest-metric CRC-valid candidate from that group as
$\hat{\mathbf{u}}_{\mathrm{PMR}}$. If no selected prefix produces a CRC-valid
candidate, the decoder returns the best-metric candidate
$\hat{\mathbf{u}}_{\mathrm{fb}}$ from the original CA-SCL final list.



\section{Simulation Setup}
\label{sec:simulation_setup}

For each code, all decoders share the code, channel,
CRC, rate profile, and initial list size. Recovery budgets
differ; average and maximum path-step costs are not matched.

\subsection{Code and Channel Parameters}

The simulations use the CRC-aided polar-coded BI-AWGN system model described in
Section~\ref{sec:system_model}. We consider two block/data-length pairs, $(N,A)=(128,64)$ and $(256,128)$. An 11-bit CRC is used. This results in polar information lengths $K=A+r$, namely $K=75$ for $N=128$ and $K=139$ for $N=256$. The CRC generator polynomial is $g(D)=D^{11}+D^{10}+D^9+D^5+1$.

For $(N,A)=(128,64)$, the information set is selected using the RM rate
profile. For $(N,A)=(256,128)$, the information set is selected using the
polarization-weight (PW) rate profile based on the beta-expansion construction
of~\cite{he2017beta}, with $\beta=2^{1/4}$.

As a construction-sensitivity check, we also formed information sets by
ranking the synthesized bit-channels according to genie-aided Monte-Carlo
estimates of their error probabilities $P_e$. At a design
$E_b/N_0=2.6$~dB, the resulting information set for $(N,K)=(128,75)$
differed from the adopted RM profile in only two positions, whereas for
$(N,K)=(256,139)$ it differed from the adopted PW profile in only one
position. For the latter case, index $196$ was replaced by index $147$.
The same substitution was obtained at design points of $2.4$ and
$2.8$~dB, indicating that the adopted rate profiles are very close to
those obtained from direct bit-channel error-probability ranking over the
considered SNR range.

For all simulations, exact channel LLRs are used at the decoder input, and
exact LLR update rules are used during decoding. Internal LLR values are
clipped to the interval $[-50,50]$ for numerical stability. The reported
$E_b/N_0$ values are normalized using the data rate $A/N=1/2$. No hidden
SNR offset is applied.

All indices are zero based. The RM profile ranks indices by decreasing
$(\operatorname{wt}(i),i)$, so equal-weight indices are resolved in favor
of the larger index. For PW, $i=\sum_j b_j2^j$, with $b_0$ the
least-significant bit, and indices are ranked by decreasing
$\sum_j b_j\beta^j$, where $\beta=2^{1/4}$. Selected information
indices are ordered increasingly. We use the non-bit-reversed
Ar{\i}kan transform with natural $u_0,\ldots,u_{N-1}$ decoding order.
The CRC is appended systematically using zero initialization, no bit
reflection, and zero final XOR. With payload rate $R_{\mathrm{data}}=A/N=1/2$, the
noise variance is
$\sigma^2=[2R_{\mathrm{data}}10^{(E_b/N_0)_{\mathrm{dB}}/10}]^{-1}$.

\subsection{Compared Decoders and Recovery Parameters}

All decoders use initial CA-SCL list size $L=8$. All recovery methods are
CRC-triggered and are activated only when the initial CA-SCL pass does not
produce a CRC-valid candidate. The baseline figures compare
CA-SCL~\cite{niu2012}, Cheng-SCLF~\cite{cheng2019bitflip},
Pan-GSCLF-1~\cite{pan2020gsclf},
SP-SCLF$(k=L/2,T=15)$~\cite{rowshan2022sclistflip},
D-SCLF-2~\cite{shen2022dynamic}, and FSP-SCL using the SSP2
path-distance variant~\cite{wang2024shift}. D-SCLF-2 uses order two,
$\alpha=0.4$, and the common tail-CRC configuration; D-SCLF-2 and
FSP-SCL use $T=15$. The D-SCLF-2 comparison concerns the specified
tail-CRC configuration rather than the distributed-CRC
configuration of the original 5G decoder.

The ablation includes SCLF-Diff with $T=15$ (the proposed
decoder's anchor) and $T=20$ (a larger-retry reference), using
\eqref{eq:sp_sclf_score}~\cite{ivanov2021gsclf} and retaining
the discarded $L$ paths per retry. The proposed decoder adds
PMR only after the $T=15$ stage fails; comparison with this
anchor measures the fallback's incremental FER benefit and
path-step cost. Costs are not matched to the $T=20$ reference.
All methods use the exact metric in~\eqref{eq:path_metric}.

Both PMR-only variants run directly after CA-SCL failure.
Fixed-attempt PMR processes up to the $T=15$ highest-ranked
prefixes with $L_j^{\mathrm{loc}}=8$; full-budget PMR uses
coverage-based selection and $Q=128$. These variants evaluate
stored-prefix recovery without SCLF-Diff.

For PMR-based methods, $B_{\max}=64$ and
$L_{\max}^{\mathrm{loc}}=8$. Full-budget PMR and the PMR fallback in the
proposed decoder use $\rho=0.99$, $M_{\min}=8$, $Q=128$, and
prefix-processing group size $R=8$.

The PMR parameters were initially selected
heuristically to limit storage and recovery complexity.
An earlier configuration used $B_{\max}=32$, $\rho=0.95$,
$Q=64$, $L_{\max}^{\mathrm{loc}}=4$, and $R=4$.
Preliminary comparisons showed weaker FER than
SCLF-Diff $(T=15)$, motivating an increase to the
reported values. The choice $M_{\min}=8$ was heuristic.
The same parameters were used for both codes and
all SNR points.

To isolate the frozen-suffix correction, a score ablation compares
the raw score $R_j^{\mathrm{raw}}=\mathrm{PM}_j^{(i_j)}$
with the adjusted score in~\eqref{eq:pmr_rank}.
The two variants use the same CA-SCL and SCLF-Diff $(T=15)$
stages and recovery budgets but are evaluated in separate
Monte Carlo runs; the adjusted-score variant is the proposed decoder.

Cheng-SCLF uses eliminated-list-mass ranking with $\alpha=1$ and $T=15$,
while Pan-GSCLF-1 uses retained-versus-eliminated log-mass ranking with
$\alpha=1.2$ and $T=15$; both retain the eliminated $L$ paths in each
retry.

\subsection{Simulation Procedure and Complexity Metric}

FER is the fraction of frames whose decoded payload differs
from the transmitted payload, including incorrect CRC-valid
outputs. For the focused comparisons of both the $(128,64)$ RM-profile and
$(256,128)$ PW-profile codes, each SNR point is simulated until at least
$3\times10^4$ frames have been processed and the error-count targets of
$2100$ CA-SCL frame errors and $225$ proposed-decoder frame errors have
been reached, or until the maximum budget of $10^7$ frames is reached.
At $3.0$ dB for the $(256,128)$ code, a fixed budget of
$10^7$ frames is used instead of error-count-based stopping,
yielding $5205$ CA-SCL frame errors and $443$ proposed-decoder
frame errors.

The Wilson $95\%$ intervals are nominal uncertainty
summaries, as their coverage has not been calibrated for the
error-count-based stopping rule. Reported FER differences,
including the score-ablation reductions, are interpreted
descriptively without formal significance claims.

Within each Monte Carlo run, the included decoders are evaluated using
common transmitted frames and noise realizations. Baseline results obtained
from separate Monte Carlo runs are normalized by their respective simulated
frame counts. Wilson $95\%$ binomial confidence intervals are computed from
the corresponding numbers of frames and frame errors. In particular, the proposed-decoder FER at $3.0$ dB for the $(256,128)$ code is
$4.43\times10^{-5}$, with a nominal $95\%$ confidence interval of approximately
$[4.04\times10^{-5},\,4.86\times10^{-5}]$.

Decoder complexity is measured using a reconstruction-aware path-step
count. A path step is counted whenever one active decoding path is
processed at one bit position. Let $P_f$ denote the path expansions
performed by all CA-SCL, SCLF-Diff, and PMR suffix-continuation stages for
frame $f$. Because PMR does not store path-specific LLR and partial-sum
states, continuing a prefix discarded at position $i_j$ requires
replaying its $i_j+1$ stored decisions. Accordingly, the complexity of
frame $f$ is
\begin{equation}
C_f
=
P_f+
\sum_{j\in\mathcal{P}_f}(i_j+1),
\label{eq:reconstruction_complexity}
\end{equation}
where $\mathcal{P}_f$ is the set of PMR prefixes actually processed
before CRC-based early stopping. The reported average is
\begin{equation}
\bar{C}
=
\frac{1}{N_{\mathrm{frames}}}
\sum_{f=1}^{N_{\mathrm{frames}}}C_f .
\end{equation}
Reconstruction-aware complexity for the proposed decoder
was measured in a separate instrumented evaluation using
the same code, channel, and recovery parameters.
The reported count includes suffix continuation and decoder-state
reconstruction, but excludes ranking arithmetic, prefix-bank
maintenance and copying, CRC-check costs, and storage requirements.
The bank is maintained during the initial CA-SCL pass even when
PMR is not activated. These counts provide a proxy for decoding
work and do not establish runtime or memory equivalence.

For completeness, worst-case complexity is also bounded. Let
$C_{\mathrm{SCL}}$ denote the path-step cost of one complete SCL pass.
A decoder performing at most $T$ shifted-pruning retries satisfies
\begin{equation}
C_{\mathrm{SP}}^{\max}
\le (T+1)C_{\mathrm{SCL}}.
\end{equation}
For the proposed decoder, $|\mathcal{S}_{\rho}|\le B_{\max}$ and
$\sum_{j\in\mathcal{S}_{\rho}}L_j^{\mathrm{loc}}\le Q$, giving the
conservative bound
\begin{equation}
C_{\mathrm{prop}}^{\max}
\le
(T+1)C_{\mathrm{SCL}}
+B_{\max}N+QN.
\end{equation}
The three terms correspond to the initial and SCLF-Diff passes,
prefix-state reconstruction, and PMR suffix continuation, respectively.
These are conservative bounds because stored
prefixes generally begin inside the codeword and CRC-based early stopping
usually prevents processing every PMR group.

For the $(128,64)$ and $(256,128)$ codes, the corresponding conservative
bounds are $38\,144$ and $75\,520$ path steps, compared with
$17\,808$ and $34\,608$ for SCLF-Diff $(T=20)$, respectively.
Across the representative simulated SNR points, the maximum observed
reconstruction-aware costs of the proposed decoder were $25\,653$ and
$49\,765$ path steps, respectively; both exceed the corresponding
SCLF-Diff $(T=20)$ bounds. Thus, small differences in average path-step
counts at higher SNRs coexist with substantially larger costs on difficult
frames. Reducing $Q$ or $B_{\max}$ would tighten the stated bound,
with the resulting FER--complexity tradeoff requiring further evaluation.

\section{Results}

\begin{figure}[!t]
    \centering
    \includegraphics[width=\columnwidth]{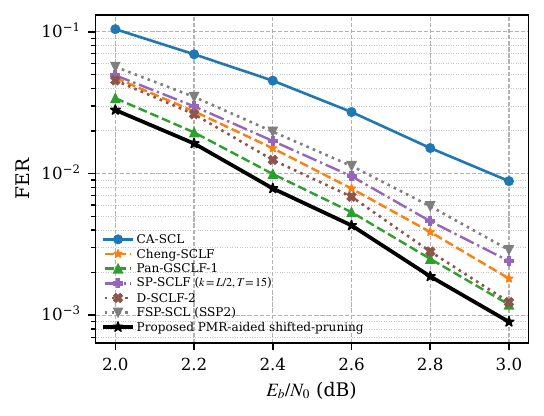}
    \caption{Baseline FER comparison for $(N,A)=(128,64)$ under the RM rate profile.}
    \label{fig:n128_baseline}
\end{figure}

\begin{figure}[!t]
    \centering
    \includegraphics[width=\columnwidth]{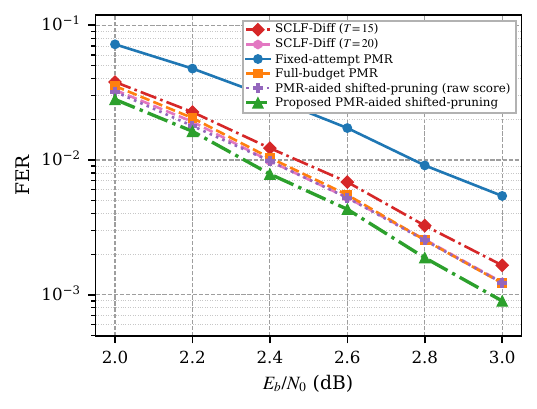}
    \caption{SCLF-Diff and PMR ablation for $(N,A)=(128,64)$ under the RM rate profile.}
    \label{fig:n128_ablation}
\end{figure}

\begin{figure}[!t]
    \centering
    \includegraphics[width=\columnwidth]{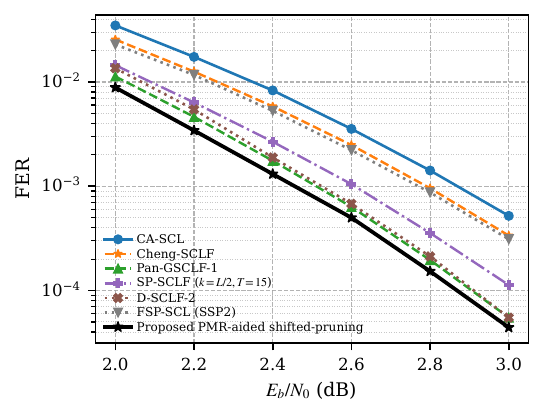}
    \caption{Baseline FER comparison for $(N,A)=(256,128)$ under the PW rate profile.}
    \label{fig:n256_baseline}
\end{figure}

\begin{figure}[!t]
    \centering
    \includegraphics[width=\columnwidth]{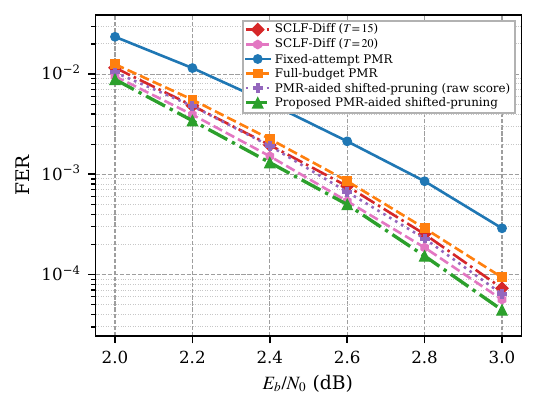}
    \caption{SCLF-Diff and PMR ablation for $(N,A)=(256,128)$ under the PW rate profile.}
    \label{fig:n256_ablation}
\end{figure}

Figs.~\ref{fig:n128_baseline}--\ref{fig:n256_ablation} show
the baseline and ablation FER comparisons for the $(128,64)$
RM-profile and $(256,128)$ PW-profile codes, using the
configurations in Section~\ref{sec:simulation_setup}.

For the $(128,64)$ code, the proposed decoder gives the lowest observed
FER among the evaluated configurations at every simulated SNR. For the $(256,128)$ code, it gives the lowest observed FER at all six points.
At $2.4$ dB, SCLF-Diff $(T=20)$ and the proposed decoder achieve FERs
of $1.515\times10^{-3}$ and $1.310\times10^{-3}$, respectively, with
overlapping nominal confidence intervals. The proposed decoder has lower observed
FER than SCLF-Diff $(T=15)$ and the considered literature baselines
under their specified budgets. Its observed FER is close to that of
SCLF-Diff $(T=20)$, although neither this comparison nor the interval
overlap establishes statistical equivalence or an advantage at equal
average complexity.

In the ablation, the adjusted-score decoder gives the lowest
observed FER among the PMR-based variants at every simulated
SNR. The estimated FER reductions relative to the raw-score
variant are $8.8\%$--$26.7\%$ for $(128,64)$ and
$15.0\%$--$33.0\%$ for $(256,128)$. This comparison uses common recovery parameters,
although the variants are evaluated in separate Monte Carlo runs.
Using the transmitted sequence only for diagnosis, the adjusted-score
bank contains a correct discarded prefix in $93.9\%$--$97.8\%$
and $87.0\%$--$92.7\%$ of CA-SCL list-miss frames for
$(128,64)$ and $(256,128)$, respectively, over $2.0$--$2.8$ dB.

Table~\ref{tab:avg_path_expansions} reports average
reconstruction-aware path steps. At $2.0$ dB, the proposed decoder has the highest average
count among the listed decoders for $(128,64)$. For
$(256,128)$, its average count exceeds those of both
SCLF-Diff configurations but is lower than those of
Cheng-SCLF and FSP-SCL.
Relative to SCLF-Diff $(T=20)$, its overhead is $9.4\%$ for
$(128,64)$ and $4.3\%$ for $(256,128)$; relative to SCLF-Diff
$(T=15)$, the corresponding overheads are $18.4\%$ and $8.0\%$.
At $2.6$ dB, the overhead relative to SCLF-Diff $(T=20)$ decreases
to $2.7\%$ and $0.3\%$, and at $3.0$ dB it is below $0.6\%$
for both codes. Reconstruction contributes $100.1$ and $74.5$ average
steps at $2.0$ dB, decreasing to $3.0$ and $0.4$ at $3.0$ dB,
respectively. Conditioned on CA-SCL failure for $(256,128)$, the
proposed decoder uses $5.7\%$--$13.8\%$ fewer complete
shifted-pruning retries than SCLF-Diff $(T=20)$. This reduction concerns complete
retries only: the PMR continuation and reconstruction costs remain
included in the total. The results characterize a FER--path-step
tradeoff under the evaluated settings; they do not establish a gain
at matched average complexity.

\begin{table}[!t]
\caption{Average Reconstruction-Aware Path Steps per Frame at
Representative SNRs under the Specified Recovery Budgets}
\label{tab:avg_path_expansions}
\centering
\scriptsize
\setlength{\tabcolsep}{1.5pt}
\resizebox{\columnwidth}{!}{%
\begin{tabular}{@{}l*{6}{c}@{}}
\toprule
& \multicolumn{3}{c}{$(128,64)$, RM}
& \multicolumn{3}{c}{$(256,128)$, PW} \\
\cmidrule(lr){2-4}\cmidrule(lr){5-7}
Decoder
& \multicolumn{3}{c}{$E_b/N_0$ (dB)}
& \multicolumn{3}{c}{$E_b/N_0$ (dB)} \\
& 2.0 & 2.6 & 3.0
& 2.0 & 2.6 & 3.0 \\
\midrule

CA-SCL
& 848.0 & 848.0 & 848.0
& 1648.0 & 1648.0 & 1648.0 \\

Cheng-SCLF
& 1728.5 & 1043.4 & 904.2
& 2374.5 & 1718.6 & 1658.1 \\

Pan-GSCLF-1
& 1543.7 & 989.1 & 885.8
& 2123.0 & 1683.0 & 1652.2 \\

SCLF-Diff $(T=15)$
& 1613.7 & 1018.5 & 896.5
& 2123.3 & 1687.6 & 1653.0 \\

SP-SCLF$(k=L/2,T=15)$
& 1665.6 & 1032.9 & 899.5
& 2162.0 & 1691.8 & 1653.6 \\

D-SCLF-2
& 1644.5 & 1006.2 & 888.6
& 2140.3 & 1684.7 & 1652.1 \\

FSP-SCL (SSP2)
& 1770.3 & 1064.9 & 911.6
& 2336.4 & 1716.5 & 1657.6 \\

SCLF-Diff $(T=20)$
& 1746.6 & 1041.3 & 901.9
& 2199.5 & 1691.1 & 1653.4 \\

\addlinespace
PMR-aided shifted-pruning
& 1910.3 & 1069.2 & 906.8
& 2293.0 & 1696.0 & 1653.8 \\

\bottomrule
\end{tabular}%
}
\end{table}

\section{Conclusion}

This paper proposed a PMR-aided shifted-pruning SCL-flip decoder that
locally continues discarded prefixes stored during the initial CA-SCL
pass after SCLF-Diff recovery fails. Its distinguishing mechanism is
the combination of stored-prefix reuse, cross-depth mass-based ranking,
and bounded local continuation in a CRC-triggered fallback stage.
For the tested codes and configurations, this fallback gives lower
observed FER than SCLF-Diff $(T=15)$, while its observed FER is
close to that of SCLF-Diff $(T=20)$. These gains involve additional
path-step cost, particularly at $2.0$ dB, and a larger worst-case bound.
The present comparisons characterize this tradeoff; establishing an
advantage over a larger-retry SCLF-Diff decoder at equal average
complexity remains open.

Extending stored-prefix recovery to list-based PAC
decoding would require storing and locally continuing
prefixes with the corresponding PAC path metric, retaining
or reconstructing their convolutional-precoder states, and
developing and validating PAC-specific ranking and
continuation rules.

\balance
\bibliographystyle{IEEEtran}
\bibliography{references}

\end{document}